\documentclass[reprint,amsmath,amssymb,aps,superscriptaddress]{revtex4-2}
\usepackage{times}
\usepackage[pdftex]{graphicx}
\usepackage{dcolumn}
\usepackage{bm}
\usepackage{amsmath}
\usepackage{indentfirst}
\usepackage{float}
\usepackage[colorlinks,linkcolor=blue,anchorcolor=blue,citecolor=blue]{hyperref}
\usepackage[dvipsnames]{xcolor}

\usepackage{braket}
\usepackage{multirow}
\usepackage{changes}

\usepackage{makecell}

\newcommand{\sgz}{\hat{\sigma}^z}
\newcommand{\sgy}{\hat{\sigma}^y}
\newcommand{\sgx}{\hat{\sigma}^x}

\newcommand{\Uop}{\mathcal{U}}
\newcommand{\PQC}{\mathcal{C}(\vec{\theta})}

\newcommand{\Real}{{\rm Re}}
\newcommand{\similarity}{{\rm similarity}}
\newcommand{\accuracy}{{\rm accuracy}}

\begin{document}

\title{Variational quantum process tomography}

\author{Shichuan Xue}
\affiliation{%
 Institute for Quantum Information \& State Key Laboratory of High Performance Computing, College of Computer Science and Technology, National University of Defense Technology, Changsha 410073, China
}%

\author{Yong Liu}%
\affiliation{%
 Institute for Quantum Information \& State Key Laboratory of High Performance Computing, College of Computer Science and Technology, National University of Defense Technology, Changsha 410073, China
}
\affiliation{%
 College of Information and Communication, National University of Defense Technology, Xi'an 710006, China
}

\author{Yang Wang}%
\affiliation{%
 Institute for Quantum Information \& State Key Laboratory of High Performance Computing, College of Computer Science and Technology, National University of Defense Technology, Changsha 410073, China
}

\author{Pingyu Zhu}%
\affiliation{%
 Institute for Quantum Information \& State Key Laboratory of High Performance Computing, College of Computer Science and Technology, National University of Defense Technology, Changsha 410073, China
}

\author{Chu Guo}
\email{guochu604b@gmail.com}
\affiliation{Key Laboratory of Low-Dimensional Quantum Structures and Quantum Control of Ministry of Education, Department of Physics and Synergetic Innovation Center for Quantum Effects and Applications, Hunan Normal University, Changsha 410081, China}

\author{Junjie Wu}
\email{junjiewu@nudt.edu.cn}
\affiliation{%
 Institute for Quantum Information \& State Key Laboratory of High Performance Computing, College of Computer Science and Technology, National University of Defense Technology, Changsha 410073, China
}%

\begin{abstract}
Quantum process tomography is an experimental technique to fully characterize an unknown quantum process. Standard quantum process tomography suffers from exponentially scaling of the number of measurements with the increasing system size. In this work, we put forward a quantum machine learning algorithm which approximately encodes the unknown unitary quantum process into a relatively shallow depth parametric quantum circuit. 
We demonstrate our method by reconstructing the unitary quantum processes resulting from the quantum Hamiltonian evolution and random quantum circuits up to $8$ qubits. Results show that those quantum processes could be reconstructed with high fidelity, while the number of input states required are at least $2$ orders of magnitude less than required by the standard quantum process tomography.

\end{abstract}

\date{\today}
\maketitle

\address{}

\vspace{8mm}
\section{\label{sec:intro}INTRODUCTION}

Quantum process tomography is an indispensable technique in quantum information processing to fully characterize an unknown quantum process\cite{Chuang2000Quantum}. 
It is increasingly pivotal in identifying and verifying the performance of a quantum device and its dynamics when the system goes larger.
Standard quantum process tomography (SQPT) works by preparing an informationally complete set of input states and then performing the standard quantum state tomography on the corresponding output quantum states~\cite{ChuangNielsen1997,ArianoPresti2001}. As a result, the total number of quantum measurements scales as $4^{2n}$ for an $n$-qubit quantum process. The exponential growth severely limits the problem size on which SQPT can be feasibly applied. Currently, SQPT has only been experimentally implemented up to $3$ qubits~\cite{Bialczak2009Quantum,O2004Quantum,Childs2000Realization,shabani2011efficient,riebe2006process,govia2020bootstrapping}. In the meantime, with the rapid development of quantum computing hardwares~\cite{guo2019general,AruteMartinisQuantumSupremacy2019,WuPan2021}, scalable quantum process tomography schemes are in great demand.

Various schemes have been proposed to alleviate the exponential scaling problem of SQPT. For example, ancilla-assisted process tomography could reduce the exponential number of input states to a single entangled state~\cite{D2001Quantum,Altepeter2003Ancilla,Hou2020Experimental}, and direct characterization of quantum dynamics (DCQD) could reduce the total number of configurations from $4^{2n}$ to $4^{n}$~\cite{Mohseni2006Direct,2007Direct}. In general, the exponential scaling is unavoidable to reconstruct a generic quantum process. However, by assuming certain structures of the unknown quantum process, the number of configurations can be significantly reduced. Such examples include compressed sensing quantum process tomography that assumes the measurement outcomes are sparse~\cite{RodionovKorotkov2014}, and tensor network states based quantum process tomography which assumes a low entanglement structure of the underlying quantum process~\cite{GuoPoletti2020,TorlaiAolita2020}. Recently,  it is shown that one could efficiently encode the information of certain quantum states into a parametric quantum circuit (PQC) using a gradient-based quantum machine learning algorithm, after which the unknown quantum state can be reconstructed classically with high fidelity using the optimal parameters of the PQC~\cite{LiuWu2020}.

In this work, we propose a unsupervised quantum machine learning algorithm for quantum process tomography, which is also a continuation of Ref.~\cite{LiuWu2020}. As shown in Fig. \ref{fig:PQC}, we use a PQC of certain depth $d$ to approximate the unknown quantum process denoted by $\Uop$, where $\vec{\theta}$ is a list of parameters to be optimized in this PQC. To learn the information of $\Uop$, we randomly prepare a set of $N$ random quantum states $\vert\psi_j\rangle$, each of which is separately fed into the unknown quantum process and the PQC. Then as long as each pair of output quantum states, $\Uop \vert\psi_j\rangle $ and $\PQC \vert\psi_j\rangle $ are equal to each other, and that $N$ is large enough, the unitary operation represented by $\PQC$ should be approximate to $\Uop$. As a result, all the information of $\Uop$ are stored in the parameters $\vec{\theta}$, and we can systematically reconstruct $\Uop$ from those parameters using a classical computer. Interestingly, during the training process, we use an additional set of random input states as the validation set, similar to that used in classical machine learning algorithms, which tests the generalization ability of the training outcomes for a particular PQC and set of input states. Here we also point out that our approach is initially designed only for the unitary quantum process, however, it is possible to generalize it to a generic quantum process by using auxiliary qubits as in Ref.~\cite{LiuWu2020}.

Compared to its alternatives, our approach has several advantages. First, near-term quantum computers or simulators may only be able to faithfully run a quantum process with limited depths, therefore it is reasonable that the same process could be reproduced by a parametric quantum circuit with a relatively low depth. However, the amount of entanglement produced by such quantum computers could be huge~\cite{AruteMartinisQuantumSupremacy2019,WuPan2021}, in which case the tensor network states based methods would be invalid. Second, our approach only requires to measure a single qubit for each configuration, taking into consideration that for current quantum computers the errors of quantum gate operations are almost one order of magnitude less than that of quantum measurements~\cite{AruteMartinisQuantumSupremacy2019,WuPan2021}, our approach could be less prone to errors. Last, with a small number of input states, our approach may already be able to reconstruct $\Uop$ with very high fidelity. We demonstrate our approach on the reconstruction of two unitary processes produced by quantum Hamiltonian evolution and random quantum circuits, respectively. In both examples our numerical results show that we could reconstruct a quantum process up to $8$ qubits with a $\similarity$ value (defined in Eq.(\ref{eq:similarity})) higher than $97\%$, and the number of required input quantum states is smaller than that required by SQPT by at least $2$ orders of magnitude.

This paper is organized as follows. In Sec.~\ref{sec:I}, we introduce the scheme of our quantum machine learning algorithm for quantum process tomography. In Sec.~\ref{sec:II}, we demonstrate our method with numerical simulations of quantum process tomography for the time evolution of a quantum \textit{XXZ} spin chain and the randomly generated quantum circuit. We conclude in Sec.~\ref{sec:III}.

\section{Approximating unitary quantum processes with parametric quantum circuit}\label{sec:I}

\begin{figure}[ht]
\includegraphics[width=\linewidth]{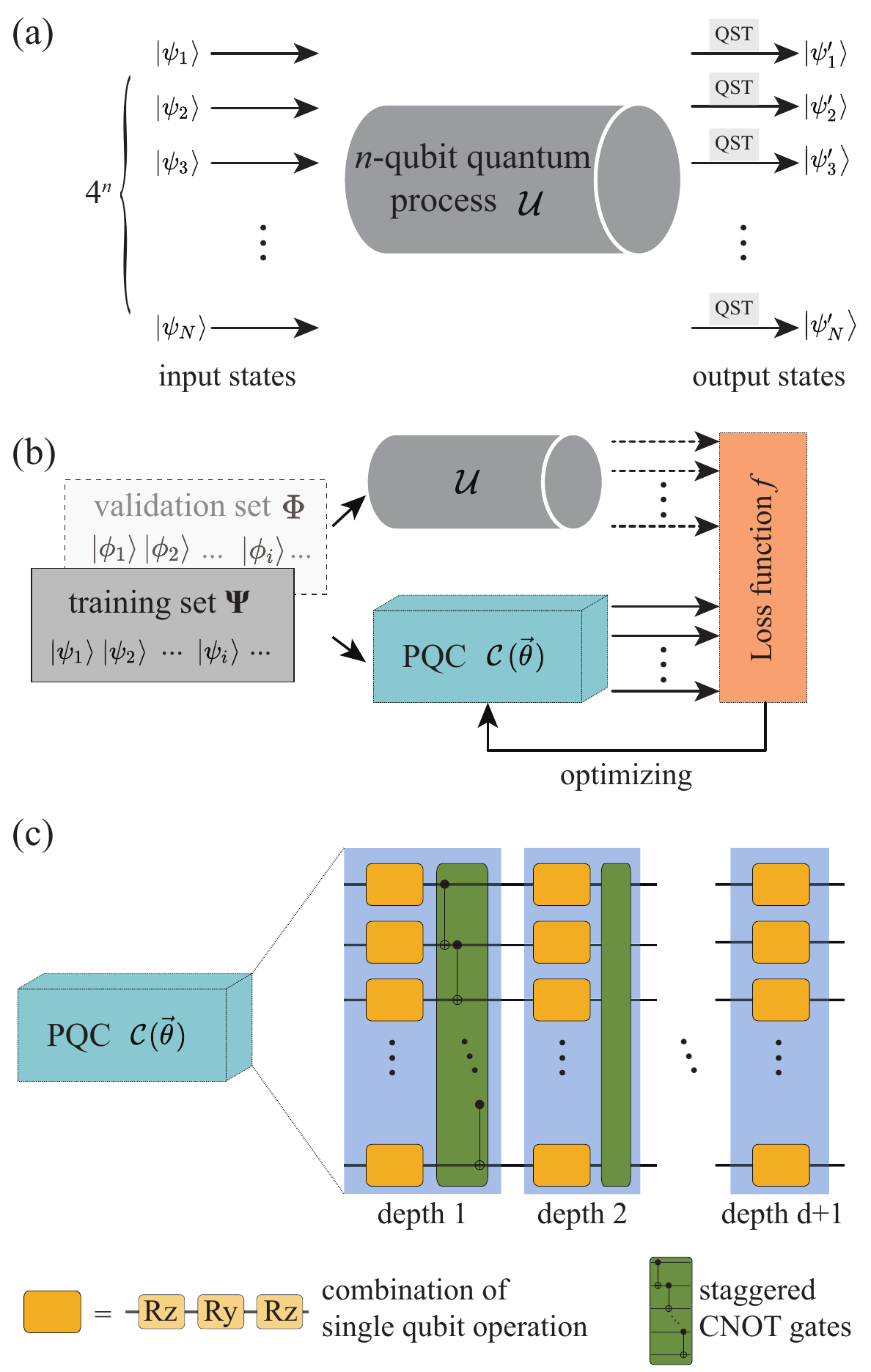}
\caption{Scheme of variational quantum process tomography. (a) illustrates the procedures of standard quantum process tomography. The information of the unknown quantum process is learned by $4^n$ pairs of inputs and outputs for an $n$-qubit quantum process $\Uop$. (b) gives a general framework of our method. We build a loss function $f$ to evaluate the distance between $\Uop$ and $\PQC$. By training the PQC with the quantum states in the training set and validation set, and  optimizing the parameters in $\PQC$ based on a gradient-descending approach, the PQC gradually approximates the physical quantum process $\Uop$. (c) shows the structure of the parametric quantum circuit which begins and ends with a single-qubit layer. Each two-qubit layer is counted as a depth and the circuit contains $d$ depths of operations and ends with a single-qubit layer. 
\label{fig:PQC} }
\end{figure}

Our quantum machine algorithm is composed of three parts: design of the PQC, training, and validation, which are shown in Fig.~\ref{fig:PQC}(b, c). Fig.~\ref{fig:PQC}(a) shows the procedures of standard quantum process tomography as a comparison. In the following, we present the details of each component.

\subsection{The parametric quantum circuit}
The design of our PQC is shown in Fig.~\ref{fig:PQC}(c), where interlaced layers of single-qubit gates and two-qubits gates are used. It is designed to quickly generate entanglement between qubits, thus make it possible to approximate complicated quantum processes. In practice, the design of the PQC should also take the underlying quantum hardware into consideration, especially the choice as well as the pattern of the two-qubit gates. Each two-qubit layer is counted as a depth. Each single-qubit layer contains three rotational gates ($R_z$, $R_y$, and $R_z$) on each qubit, where $R_y$ and $R_z$ are defined as
\begin{align}
R_{y}(\theta)=&\left[\begin{array}{rr}
\cos \frac{\theta}{2} & -\sin \frac{\theta}{2} \vspace{0.5ex}\\
\sin \frac{\theta}{2} & \cos \frac{\theta}{2}
\end{array}\right]; \\ 
R_{z}(\theta)=&\left[\begin{array}{cc}
e^{ -i \frac{\theta}{2}} & 0 \\
0 & e^{i \frac{\theta}{2}}
\end{array}\right].
\end{align}
The sequence $R_z$, $R_y$, and $R_z$ makes sure that arbitrary single qubit rotations can be produced with appropriate parameters. Our PQC ends with a single-qubit layer. As a result, for such a circuit with $n$ qubits and $d$ depth, the total number of parameters is $3n(d + 1)$.

\subsection{The training process}
We build a loss function which reflects the distance between the unitary represented by $\PQC$ and the target unitary $\Uop$. Concretely, we first randomly generate a set of $N$ quantum states, denoted as $\Psi= \left\{\ket{\psi_1},\ket{\psi_2},...,\ket{\psi_N}\right\}$. Here, a random quantum state is generated by applying an $R_y$ gate with random parameters onto each qubit and $CZ$ gates with random control and target qubits. Each state $\ket{\psi_j}$ is fed into the unknown quantum process $\Uop$ and the PQC $\PQC$, with the output quantum states denoted as $ \ket{\psi^{ideal}_j} = \Uop\ket{\psi_j}$ and $ \ket{\psi^{\prime}_j} = \PQC\ket{\psi_j}$. Then we compute the Euclidean distance between $\ket{\psi^{ideal}_j}$ and $\ket{\psi^{\prime}_j}$, which is
\begin{align}\label{eq:distance}
\|\ket{\psi_j^{\prime}}-\ket{\psi_j^{ideal}}\|^{2} = 2 - 2 \Real\left(\braket{\psi_j^{ideal}|\psi_j^{\prime}} \right).
\end{align}
The inner product on the right-hand side of Eq.(\ref{eq:distance}) can be efficiently computed with a quantum computer using a generalized SWAP-test algorithm, which is detailed in Appendix.~\ref{app:gswap}. The loss function $f$ is defined as the summation of the distance obtained over all input states, which is
\begin{align}\label{eq:loss}
f(\vec{\theta})&=\frac{1}{N} \sum_{j=1}^{N}\|\ket{\psi_j^{\prime}}-\ket{\psi_j^{ideal}}\|^{2}\nonumber\\
&=\frac{2}{N}\sum_{j=1}^{N}\left(1-\Real\left(\braket{\psi_j^{theory}|\psi_j^{\prime}}\right)\right),
\end{align}
namely, $f$ is the mean square error between the two set of output quantum states.

The loss function $f$ is a hybrid quantum-classical function, where it contains functions to be evaluated with a quantum computer (SWAP-test) and functions to be evaluated on a classical computer. The gradient of the quantum part can be computed by 
\begin{equation} \label{eq:qgrad}
\frac{\partial \mathcal{O}(\vec{\theta})}{\partial \theta_{j}}=\frac{1}{2} \mathcal{O}\left(\vec{\theta}_{j}^{+}\right)-\frac{1}{2} \mathcal{O}\left(\vec{\theta}_{j}^{-}\right),
\end{equation}
where $\mathcal{O}$ in our case means the SWAP-test with parametric quantum gates, $\vec{\theta}_j$ denotes the $j$-th parameter in the parameter list $\vec{\theta}$, and $\vec{\theta}_j^{\pm} = \vec{\theta}_j \pm \frac{\pi}{2}$. The gradient of the loss function $f$ can be computed following Ref.~\cite{LiuHuang2021}, which proposes a method to embed Eq.(\ref{eq:qgrad}) into the classical automatic differentiation framework, such that the gradient of a hybrid quantum-classical loss function can be automatically computed using a hybrid quantum-classical computer. The gradient can then be fed into a gradient-based optimizer to minimize the loss function $f$.

After the training, we evaluate the accuracy between $\Uop$ and $\PQC$ using the $\similarity$ defined as
\begin{align}\label{eq:similarity}
\similarity(\Uop, \PQC) = 1 - \frac{\| \PQC - \Uop \|}{2\|\ \Uop \|}.
\end{align}
Here $\| X \|$ denotes the $2$-norm of the vectorized matrix $X$. $\similarity(\Uop, \PQC)=1$ means $\| \PQC - \Uop \|=0$, in which case $\Uop$ can be perfected reconstructed from $\PQC$.

\subsection{Usage of a validation set}
As a prior, we do not know whether our PQC is expressive enough or not to represent $\Uop$, and whether the number of input states is enough or not to ensure convergence to $\Uop$. Moreover, in practice we may also have the problem of overfitting such that the optimal $\PQC$ is very distinct from $\Uop$ but the loss function $f$ has already converged to $0$. To overcome these problems, we borrow the idea of the validation set from classical machine learning, which is part of the training data and primarily used to test the generalization ability of the training outcomes without resorting to the testing data.

Concretely, we generate another set of input states denoted as $\Phi= \left\{\ket{\phi_1},\ket{\phi_2},...,\ket{\phi_N}\right\}$,  i,e, the validation set, which is independent of the training set. After the training process, we feed each $\ket{\phi_j}$ into the unknown quantum process and the resulting optimal PQC, obtaining two outputs $\vert \phi_j^{ideal}\rangle = \Uop \ket{\phi_j}$ and $\ket{\phi_j^{\prime} } = \PQC \ket{\phi_j}$. Then we compute the quantum fidelity between $\ket{\phi_j^{ideal}}$ and $\ket{\phi_j^{\prime} } $ efficiently through the SWAP-test~\cite{BuhrmanWolf2001} on a quantum computer, and summarize over all the instances of the validation set, which is defined as the $\accuracy$
\begin{align}\label{eq:accuracy}
\accuracy(\Uop, \PQC)=\frac{1}{N}\sum_{j \in \Phi}^{N}\left|\braket{\phi_j^{ideal}|\phi_j^{\prime}}\right|.
\end{align}
We can see that $0\leq \accuracy(\Uop, \PQC) \leq 1$. If $\accuracy$ is close to $1$, it means the PQC we obtained can be well generalized to the new input state. Therefore, it is possible to determine if the training is successful based on $\accuracy$, without resorting to the truth value of $\Uop$ as required in Eq.(\ref{eq:similarity}). In our numerical simulation we show that $\accuracy$ and $\similarity$ are indeed strongly correlated, therefore we can identify the simulation with $\accuracy$ close to one as the most faithful reconstruction of $\Uop$.

\section{Numerical results and discussions} \label{sec:II}

We demonstrate our quantum machine learning algorithm using numerical simulations based on a classical PQC simulator. Specifically, we concentrate on two cases: 1) the unknown unitary process is produced by a quantum Hamiltonian evolution and 2) by random quantum circuits, respectively.

\subsection{\textit{XXZ} spin chain time evolution}
We take the Hamiltonian of the Heisenberg \textit{XXZ} spin chain \cite{Marko2008Many} in a magnetic field as our example, which can be written as, 
\begin{equation}
\hat{H}_{\rm XXZ}=\sum_{l=1}^{n-1}\left[
J\left(\sgx_l \sgx_{l+1} +\sgy_l \sgy_{l+1}\right)+
\Delta \sgz_l\sgz_{l+1} \right]+
h \sum_{l=1}^{n} \sgz_l.
\end{equation}
Here $n$ is the number of spins (qubits), $J$ is the tunneling strength, $\Delta$ is the interaction strength, and $h$ is the magnetization strength. The evolutionary operator with time $dt$ is denoted as 
\begin{align}
\Uop_{{\rm XXZ}}=e^{-i\hat{H}_{\rm {XXZ}}dt}.
\end{align}
In the simulations, we fix $h=0.1$, $J=1$, and set $dt=0.01$. 

\begin{figure*}[ht]
\includegraphics[width=\linewidth]{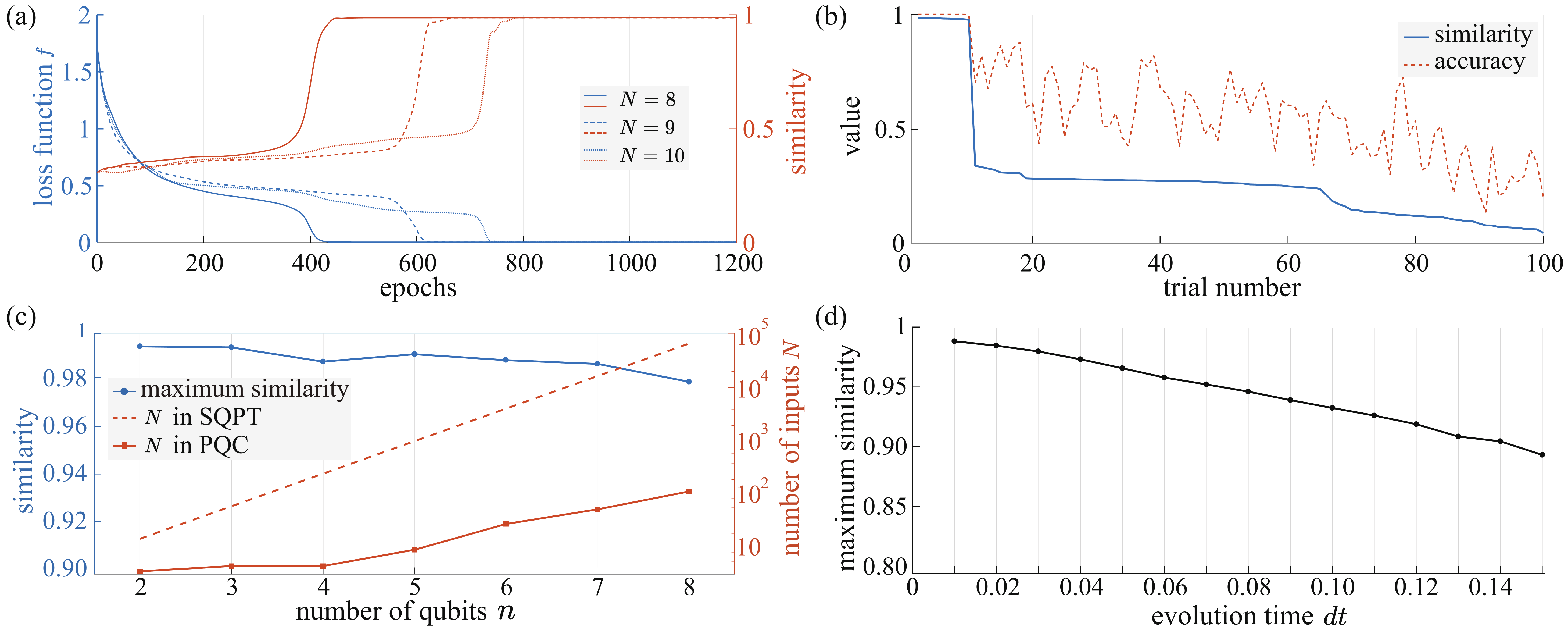}
\caption{\label{fig:xxz} Numerical results on the Heisenberg \textit{XXZ} spin chain time evolution. (a) shows a typical training process on $(5,6,8)$, $(5,6,9)$ and $(5,6,10)$ configurations. The loss function $f$ (blue line, corresponds to the left axis) decreases with the training epochs till converging to a threshold value. Meanwhile, the $\similarity$ value (orange line, corresponds to the right axis) approaching $1$. (b) shows the correspondence between the $\accuracy$ on validation set and the $\similarity$ of the final results among 100 trials of $(5,6,10)$ simulations. The data is sorted in a descending $\similarity$ order. The blue solid line denotes the $\similarity$ between $\PQC$ and $\Uop$. The red dotted line shows the $\accuracy$ value on the validation set. It is noted that the $\accuracy$ and $\similarity$ values are indeed strongly correlated. (c) Scalability tests from $2$-qubit to $8$-qubit quantum processes. The left-axis (blue circle) shows the maximum $\similarity$ reached during repetitive trials and the logarithmic right-axis represents the corresponding number of input states $N$ needed. The orange dotted line denotes the number of inputs in SQPT, and the orange rectangular represents that in PQC. In (a), (b) and (c), the evolution time $dt=0.01$. (d) Extension tests on quantum processes with longer evolution time $dt$, ranging from $0.01$ to $0.15$. The line graph shows the maximum $\similarity$ value reached among repetitive trials on a $(6,6,30)$ circuit.}
\end{figure*}

As shown in Sec.  \ref{sec:I}, we will encode $\Uop_{{\rm XXZ}}$ into the circuit parameters $\vec{\theta}$. We prepare a set of randomly generated states $\ket{\psi_{i}}$ as inputs, i.e., the training set $\Psi$, and feed each $\ket{\psi_i}$ into both the quantum process and the PQC. The same training procedure is repeated for $100$ times, with each of them initialized independently. After training, another randomly generated validation set 
is utilized to test the generalizability and pick out the most faithful instance of parametric circuit. Based on Eq.(\ref{eq:similarity}) and (\ref{eq:accuracy}), we can evaluate the performance of the parametric quantum circuits on the training set and validation set.

Here, we denote $(n,d,N)$ as an $n$-qubit, $d$-depth, and $N$-input PQC configuration in numerical simulation. Fig.  \ref{fig:xxz}(a) illustrates a typical $5$-qubit training process on different sizes of input states---$(5,6,8)$, $(5,6,9)$ and $(5,6,10)$ cases. 
The parametric quantum circuit is initialized by randomly generated parameters and evaluated by loss function in Eq.(\ref{eq:loss}) at each epoch. 
Fig.  \ref{fig:xxz}(a) shows that the loss function goes down with the training process while the $\similarity$ gradually reaches $1$ under different $N$. Moreover, among $100$ independent trials with random initialization on the $(5,6,10)$ PQC configuration, we utilize the validation set to distinguish the more faithful PQC. It is plotted in Fig.  \ref{fig:xxz}(b) that the $\accuracy$ on the validation set shows a strong correspondence with the final $\similarity$ value. Therefore, it is feasible to utilize the $\accuracy$ value as a criterion to determine the optimal circuit parameters to reconstruct $\Uop$ with higher $\similarity$. 


In Fig.  \ref{fig:xxz}(c), we conduct simulations on Heisenberg \textit{XXZ} spin chains of different lengths to evaluate the scalability of our method. For an $n$-qubit Heisenberg \textit{XXZ} spin chain time evolution process ($n$ ranges from $2$ to $8$), we utilize different depths of PQC $d$, choose different sizes of training sets $N$, 
repeat such independently identical simulations for $100$ times, 
and calculate the maximum $\similarity$ achieved. Compared with standard quantum process tomography, which needs $4^n$ pairs of input states and output states, the number of required input quantum states in our method is at least $2$ orders of magnitude less (concretely $N=56$ and $N=120$ for $7$-qubit and $8$-qubit quantum processes, while the corresponding numbers in SQPT are $16384$ and $65536$, respectively). Meanwhile we utilize a relatively shallow depth circuit ($d \leq 8$) and achieve a faithful $\similarity$ (higher than $97\%$) over all cases. More data and configuration details can be seen in Appendix.~\ref{app:scalability}. Besides, our method only involves single qubit measurements for each configuration, instead of measuring in the complete set of computational basis.

We further study the influence of the evolution time $dt$ of the Hamiltonian on the results of our method in Fig.  \ref{fig:xxz}(d). We take a $(6,6,30)$ PQC configuration as an example, and set various evolution times $dt $, ranging from $0.01$ to $0.15$. Under each case, we repeat the numerical simulations and get the maximum $\similarity$ value. In Fig. \ref{fig:xxz}(d), we plot the maximum $\similarity$ against the evolution times $dt$. It can be seen that our method still achieves an acceptable $\similarity$ when the evolution time goes longer.

\subsection{Randomly generated quantum circuit}

\begin{figure}[h]
\includegraphics[width=\linewidth]{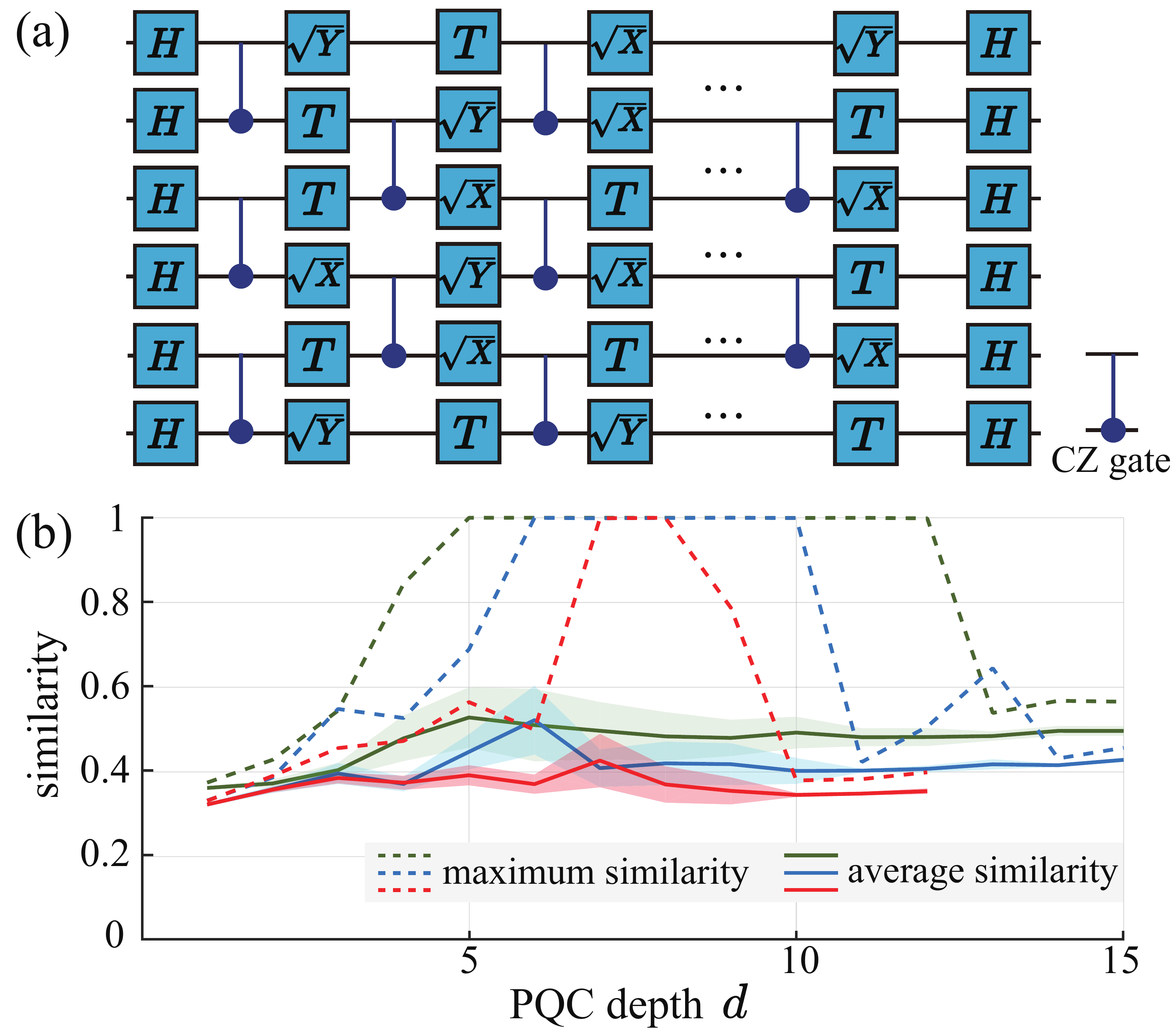}
\caption{\label{fig:rqc} Numerical results on the randomly generated quantum circuits. (a) a typical $6$-qubit randomly generated quantum circuit organization. (b) variational quantum process tomography on randomly generated quantum circuit on $4$-qubit(green), $5$-qubit(blue), and $6$-qubit(red) case, respectively. The line graph shows the relationship between the $\similarity$ value (dotted line represents maximum value, solid line shows average value) and the PQC depth, and the shaded area denotes the standard deviation.}
\end{figure}

In this section, we consider the quantum process of randomly generated circuits. For an $n$-qubit $D$-depth randomly generated quantum circuit, we firstly apply Hadamard gates to initialize the state to a symmetric superposition. Then, the circuit is organized by depth, including controlled-phase (CZ) gates alternating between odd and even configurations to entangle neighbouring qubits and randomly chosen single-qubit gate ($T$, $\sqrt{X}$ or $\sqrt{Y}$). Finally, Hadamard gates are applied to each qubit. A specific $6$-qubit randomly generated circuit is organized as shown in Fig.  \ref{fig:rqc}(a).
It is noted that such randomly generated quantum circuits are hard for efficient simulation on a classical computer \cite{boixo2018characterizing,bouland2019complexity}.

Here, we take the $4$-qubit $8$-depth, $5$-qubit $7$-depth and $6$-qubit $6$-depth randomly generated quantum circuits as examples. We utilize $(4,5,8)$, $(5,7,20)$, and $(6,8,39)$ PQC configurations to learn the corresponding quantum process respectively. Data details can be seen in Appendix.~\ref{app:scalability}.
In Fig. \ref{fig:rqc}(b), we plot the $\similarity$ value against the PQC depth $d$. It is noted that for such a specific random quantum circuit, our method can give an approximate circuit $\PQC$ with high $\similarity$ among repetitive independent trails, and the results depends on the depth of the parametric circuits $d$. Taking the $4$-qubit (green line) randomly generated quantum circuit as an example, the most appropriate depth $d$ is among $5$ to $10$. Shallow circuits may not be expressive enough to rebuild $\Uop$, while deeper depth with more parameters may result in overfitting. Hence the maximum $\similarity$ is tougher to reach one when the depth $d>12$.

\section{\label{sec:III}CONCLUSION}

In this work, we propose a quantum machine learning algorithm for quantum process tomography, which encodes the unknown
unitary quantum process into a parametric quantum circuit of certain depth $d$. A set of randomly generated product quantum states are used as the training data to minimize the loss function. The training process is repeated with a validation set in the end of filter out the instance with highest $\similarity$, namely the closest to the unknown quantum process.
We demonstrated our method by two numerical examples, including the Hamiltonian evolution with the Heisenberg \textit{XXZ} spin chain from $2$-qubit to $8$-qubit and random quantum circuits. The results indicate that a faithful reconstruction of $\Uop$ ($\similarity$ higher than $97\%$) can be reached with a relatively low-depth PQC ($d \leq 8$), and a relatively small number of training states (at least 2 orders of magnitude compared to SQPT). Moreover, only single qubit measurement is required in each configuration, instead of measuring in the complete set of computational basis. Our work presents a promising application of using quantum machine learning algorithm to accelerate quantum process tomography.

\section{ACKNOWLEDGMENTS}

We appreciate the helpful discussion with other members of the QUANTA group. We thank He-Liang Huang for fruitful discussion. J.W. acknowledges the support from the National Natural Science Foundation of China under Grant No. 62061136011 and No. 61632021. C.G. acknowledges support from National Natural Science Foundation of China through Grant No. 11805279, No. 61833010, No. 12074117 and No. 12061131011. 
\appendix

\section{generalized SWAP-test circuit} \label{app:gswap}

\begin{figure}[H]
\includegraphics[width=\linewidth]{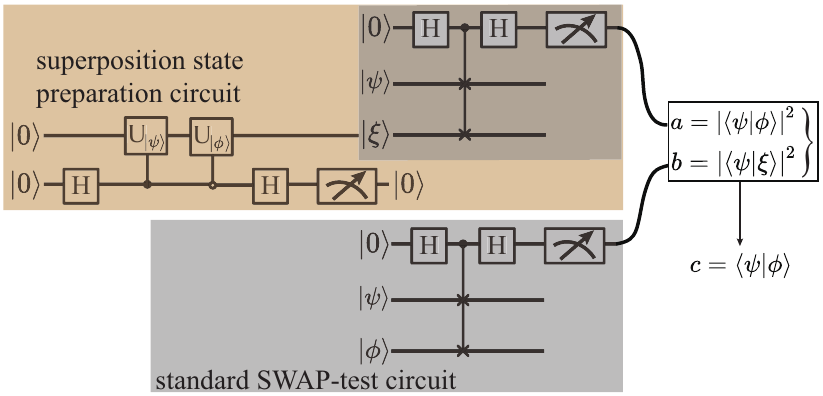}
\caption{\label{fig:swap}Generalised SWAP-test circuit. The bottom grey shaded circuit is the standard SWAP-test circuit. By measuring the ancillary qubit, we could get access to the fidelity between two input quantum states. The top orange shaded layer involves in a superposition state preparation circuit and a standard SWAP-test circuit. Solving the equations right, we can get two complex state vector's overlap value $c$.}
\end{figure}

The key to calculating the loss function and $\accuracy$ is the inner product between the target state $ \ket{\psi^{ideal}_j} = \Uop\ket{\psi_j}$ and the circuit output state $ \ket{\psi^{\prime}_j} = \PQC\ket{\psi_j}$. For real state vectors, the standard SWAP-test circuit (shown in Fig.  \ref{fig:swap} grey shaded box) is enough to evaluate the overlap since it is a real number. However, $\ket{\psi_j^{ideal}}$ and $\ket{\psi_j^{\prime}}$ are complex vectors, so we introduce a generalized SWAP-test to evaluate the overlap value.

Given two complex quantum states $\ket{\psi}$ and $\ket{\phi}$, it is already known that the fidelity between the two states can be evaluated using SWAP-test circuit, namely $a=|\braket{\psi|\phi}|^2$ can be efficiently calculated on a quantum device. Thus as long as we can prepare another superposition state $\xi=\frac{1}{\sqrt{2}}\left(\ket{\psi}+\ket{\phi}\right)$, we can also obtain the fidelity $b=|\braket{\psi|\xi}|^2$. Based on the two results above, we can arrive at the overlap between the two quantum state vectors
\begin{equation}
c=\braket{\psi|\phi}=b-\frac{a+1}{2}+i\sqrt{(a+1)b-b^2-\frac{(a-1)^2}{4}}.
\end{equation}

The superposition state $\xi$ can be easily obtained using a controlled operation with an auxiliary qubit, as $\ket{0}\bra{0}\hat{\mathcal{U}}_{\ket{\phi}}+\ket{1}\bra{1} \hat{\mathcal{U}}_{\ket{\psi}}$, where $\hat{\mathcal{U}}_{\ket{\psi}}$ denotes the unitary operation to produce the quantum state $\ket{\psi}$. As shown in Ref. \cite{zhou2011adding}, we can add control to arbitrary unitary process. By post-selecting the ancillary qubit on $\ket{0}$, we prepare the superposition state.

It is noted that our generalised approach only requires to measure a single qubit for each configuration (while the three-qubit TOFFOLI gate is used as required by SWAP test). Since the errors of quantum gate operations are almost one order of magnitude less than that of quantum measurements for current quantum computers~\cite{AruteMartinisQuantumSupremacy2019,WuPan2021}, our method could be less prone to errors, compared to massive and general measurements involved in SQPT.

\section{data details of numerical simulations} 
\label{app:scalability}

\begin{table}[h]
\caption{\label{tab:scalability} Numerical simulation details of scalability test on the Heisenberg \textit{XXZ} spin chain time evolution}
\begin{tabular}{p{1cm}<{\centering}|p{1cm}<{\centering}|p{1cm}<{\centering}|p{1.2cm}<{\centering}|p{2.2cm}<{\centering}}

$n$ & $d$ & $N_{PQC}  $ & $N_{SQPT}   $ & max. $\similarity$ \\
\hline
$2$ & $2$ & $4  $ & $16   $ & $99.46\%        $ \\
$3$ & $3$ & $5  $ & $64   $ & $99.42\%        $ \\
$4$ & $4$ & $5  $ & $256  $ & $98.82\%        $ \\
$5$ & $6$ & $10 $ & $1024 $ & $99.13\%        $ \\
$6$ & $6$ & $30 $ & $4096 $ & $98.88\%        $ \\
$7$ & $8$ & $56 $ & $16384$ & $98.73\%        $ \\
$8$ & $8$ & $120$ & $65536$ & $97.97\%      $\\
\hline
\end{tabular}
\end{table}

In Fig. \ref{fig:xxz}(c), we conduct the scalability tests on $n$-qubit Heisenberg \textit{XXZ} spin chain time evolution. Here, we give detailed numerical results listed in Table \ref{tab:scalability} where $d$, $N_{PQC}$, and $N_{SQPT}$ denote PQC depth, number of input quantum states in our method and number of input quantum states in SQPT respectively.

\begin{table}[H]
\centering
\caption{\label{tab:time} Numerical simulation details of time extension test on $(6,6,30)$ PQC configuration}
\begin{tabular}{p{1cm}<{\centering}|p{2.2cm}<{\centering}|p{1cm}<{\centering}|p{2.2cm}<{\centering}}
$dt$   &  max. $\similarity$ & $dt$   &  max. $\similarity$ \\
\hline
$0.02$ & $98.52\%$        & $0.09$ & $94.08\%$        \\
$0.03$ & $98.05\%$        & $0.10$ & $93.44\%$        \\
$0.04$ & $97.41\%$        & $0.11$ & $92.82\%$        \\
$0.05$ & $96.68\%$        & $0.12$ & $92.11\%$        \\
$0.06$ & $95.92\%$        & $0.13$ & $91.11\%$        \\
$0.07$ & $95.36\%$        & $0.14$ & $90.72\%$        \\
$0.08$ & $94.77\%$        & $0.15$ & $89.61\%$       \\
\hline
\end{tabular}
\end{table}

In Fig. \ref{fig:xxz}(d), we conduct extension tests on longer time evolution process with $(6,6,30)$ configuration. Here, we give detailed numerical results listed in Table \ref{tab:time}, where maximum and average $\similarity$ are calculated under 100 trials.

In Fig. \ref{fig:rqc}(b), we utilize $(4,5,8)$, $(5,7,20)$, and $(6,8,39)$ PQC configurations to learn the corresponding 4-qubit 8-depth, 5-qubit 7-depth and 6-qubit 6-depth randomly generated quantum circuits respectively. Here, we give detailed numerical results listed in Table \ref{tab:rqc}, where $\rm std$ corresponds to the standard deviation of 100 trials.

\begin{table*}[ht]
\caption{\label{tab:rqc} Numerical simulation details of random quantum circuit tests}

\begin{tabular}{p{0.7cm}<{\centering}|p{2cm}<{\centering}|p{2cm}<{\centering}|p{1cm}<{\centering}|p{2cm}<{\centering}|p{2cm}<{\centering}|p{1cm}<{\centering}|p{2cm}<{\centering}|p{2cm}<{\centering}|p{1cm}<{\centering}}
          & \multicolumn{3}{c|}{$4$-qubit}                & \multicolumn{3}{c|}{$5$-qubit}                & \multicolumn{3}{c}{$6$-qubit}                \\
\hline
$d$ & avg. $\similarity$ & max. $\similarity$ & std   & avg. $\similarity$ & max. $\similarity$ & std   & avg. $\similarity$ & max. $\similarity$ & std   \\
\hline
$1 $        & $36.23\%$         & $37.48\% $        & $0.008$ & $32.31\%$         & $33.26\%$         & $0.007$ & $32.31\%$         & $33.19\%$         & $0.005$ \\
$2 $        & $37.29\%$         & $42.87\% $        & $0.019$ & $35.87\%$         & $38.70\%$         & $0.017$ & $35.83\%$         & $39.04\%$         & $0.013$ \\
$3 $        & $40.43\%$         & $54.48\% $        & $0.035$ & $39.64\%$         & $54.86\%$         & $0.050$ & $38.59\%$         & $45.63\%$         & $0.026$ \\
$4 $        & $47.93\%$         & $84.37\% $        & $0.108$ & $37.18\%$         & $52.73\%$         & $0.035$ & $37.46\%$         & $47.31\%$         & $0.033$ \\
$5 $        & $52.86\%$         & $99.99\% $        & $0.143$ & $44.75\%$         & $68.94\%$         & $0.084$ & $39.22\%$         & $56.48\%$         & $0.048$ \\
$6 $        & $51.07\%$         & $99.99\%$        & $0.171$ & $52.23\%$         & $99.97\%$         & $0.163$ & $37.13\%$         & $49.97\%$         & $0.045$ \\
$7 $        & $49.71\%$         & $99.99\% $        & $0.136$ & $40.91\%$         & $99.86\%$         & $0.089$ & $42.69\%$         & $99.87\%$         & $0.126$ \\
$8 $        & $48.40\%$         & $99.99\%$        & $0.115$ & $42.03\%$         & $99.93\%$         & $0.103$ & $37.05\%$         & $99.98\%$         & $0.086$ \\
$9 $        & $48.04\%$         & $99.99\%$        & $0.086$ & $41.83\%$         & $99.98\%$         & $0.099$ & $35.54\%$         & $78.77\%$         & $0.064$ \\
$10$        & $49.29\%$         & $99.94\% $        & $0.075$ & $40.25\%$         & $99.93\%$         & $0.061$ & $34.58\%$         & $38.00\%$         & $0.010$ \\
$11$        & $48.20\%$         & $99.98\% $        & $0.042$ & $40.31\%$         & $42.36\%$         & $0.009$ & $34.90\%$         & $38.31\%$         & $0.008$ \\
$12$        & $48.24\%$         & $99.85\% $        & $0.042$ & $40.73\%$         & $50.69\%$         & $0.016$ & $35.46\%$         & $39.92\%$         & $0.014$ \\
$13$        & $48.48\%$         & $53.92\% $        & $0.022$ & $41.81\%$         & $64.49\%$         & $0.024$ &                 &          &  \\
$14$        & $49.69\%$         & $56.80\% $        & $0.024$ & $41.59\%$         & $43.19\%$         & $0.007$ &                 &          &  \\
$15$        & $49.68\%$         & $56.58\% $        & $0.023$ & $42.86\%$         & $45.70\%$         & $0.010$ &                 &          & \\
\hline
\end{tabular}
\end{table*}

\bibliographystyle{apsrev4-1}
\bibliography{refs}

\end{document}